\documentclass[12pt,twoside]{article}
\usepackage[T1]{fontenc}
\usepackage[latin1]{inputenc}
\usepackage{times}
\usepackage{graphicx}
\usepackage{a4wide}
\usepackage{listings}
 \usepackage{amsmath}

\begin{document}

\section*{A versatile method for simulating {\boldmath$pp \to ppe^+e^-$}
  and {\boldmath$dp \to pne^+e^-p_{\rm spec}$} reactions}

\thispagestyle{empty}

\begin{raggedright}

\markboth{I.~Fr\"{o}hlich {\it et al.}}
{A versatile method for simulating $pp \to ppe^+e^-$
  and $dp \to pne^+e^-p_{\rm spec}$ reactions}

I.~Fr\"{o}hlich$^{2}$, 
F.~Dohrmann$^{1}$,
T.~Galatyuk$^{2}$,
R.~Holzmann$^{3}$, 
P.K.~K\"{a}hlitz$^{1}$, 
B.~K\"{a}mpfer$^{1}$,
E.~Morini\`{e}re$^{4}$, 
Y.C.~Pachmayer$^{2}$,
B.~Ramstein$^{4}$, 
P.~Salabura$^{3,5}$, 
J.~Stroth$^{2,3}$,
R.~Trebacz$^{5}$,
J.~Van~de~Wiele$^{4}$,
and 
J.~W\"{u}stenfeld$^{1}$

\vspace{1cm}
\hspace{-0.4cm}\makebox[0.3cm][r]{$^{1}$}
Institut f\"{u}r Strahlenphysik, Forschungszentrum Dresden-Rossendorf, 01314~Dresden, Germany\\
\hspace{-0.4cm}\makebox[0.3cm][r]{$^{2}$}
Institut f\"{u}r Kernphysik, Goethe-Universit\"{a}t, 60438 ~Frankfurt, Germany\\
\hspace{-0.4cm}\makebox[0.3cm][r]{$^{3}$}
GSI Helmholtzzentrum f\"{u}r Schwerionenforschung GmbH, 64291~Darmstadt, Germany\\
\hspace{-0.4cm}\makebox[0.3cm][r]{$^{4}$}
Institut de Physique Nucl\'{e}aire d'Orsay, CNRS/IN2P3, 91406~Orsay Cedex, France\\
\hspace{-0.4cm}\makebox[0.3cm][r]{$^{5}$}
Smoluchowski Institute of Physics, Jagiellonian University of Cracow, 30-059~Krak\'{o}w, Poland

\end{raggedright}
\vspace{1cm}

\begin{center}
\textbf{Abstract}
\vspace{0.5cm}\\
\begin{minipage}[c]{0.9\columnwidth}
We have developed a versatile software package for the
  simulation of di-electron production in $pp$ and $dp$ collisions at
  SIS energies. Particular attention has been paid to incorporate
  different descriptions of the Dalitz decay $\Delta \to N e^+e^-$ via
  a common interface. In addition, suitable parameterizations for the
  virtual bremsstrahlung process $NN \to NN e^+e^-$ based on one-boson
  exchange models have been implemented.  Such simulation tools with
  high flexibility of the framework are important for the
  interpretation of the di-electron data taken with the HADES
  spectrometer and the design of forthcoming experiments.
\end{minipage}
\end{center}

\sloppy
\section{Introduction}\label{sec:intro}

Experiments with the High Acceptance Di-Electron Spectrometer (HADES)~\cite{nim}
are aimed at searching for medium modifications of hadrons at high
density and moderate temperatures created in heavy-ion
collisions in the 1-2 AGeV impact beam energy range. 
There exist a multitude of predictions, partially
conflicting in details, awaiting verification or
falsification~\cite{Schmidt08,elena}.  Due to negligible final-state
interactions with nuclear matter, di-electrons or di-muons are
considered to be  useful penetrating probes for this purpose.

While at higher energies various experimental set-ups for di-lepton measurements
have been or are operating, 
{\it e.g.}, HELIOS-3, CERES/NA35, NA38, NA50, NA60 (the latter three for di-muons), 
and PHENIX, HADES is the only presently active installation in the relativistic regime.
In addition, it can cover elementary hadron reactions 
($pp$, $\pi p$, and via tagging a spectator, also $pn$) and 
hadron--nucleus ($pA$, $\pi A$)  collisions. 
This large range of reactions is related 
to the capabilities of the SchwerIonen-Synchrotron SIS18 at GSI, Darmstadt.

When searching for medium modifications of hadrons in the di-electron
channel, it is important to have a reliable
experimental reference, in particular from
elementary hadronic reactions.  This became clearly evident in view
of the unexplained pair excess measured by the DLS and, more recently,
HADES experiments~\cite{prl,plb}.  
In addition, the knowledge of the elementary process 
$NN \to NN e^+e^-$ 
is a
prerequisite to understand possible in-medium effects in heavy-ion
dilepton data~\cite{elena,mosel1}.  
In this context, HADES has performed
two di-electron experiments using a liquid hydrogen target and
proton/deuteron beams~\cite{pp_pn,next_prl}: $pp$ at 1.25 GeV and
$dp$ at 1.25 AGeV, {\it i.e.}, at the same kinetic beam energy per
nucleon, which access a broad range of topics. For example,
the branching ratio and involved electromagnetic transition form factors  of
the $\Delta$ Dalitz decay ($\Delta \to N 
\gamma^* \to N e^+ e^-$) 
are unmeasured. In particular, the di-electron production in the
$NN$ collision is regarded to be sensitive to the nucleon form factor in the
time-like region~\cite{mosel2,jong}. 
Moreover, the cross section of
non-resonant virtual photon emission (often referred to as
``bremsstrahlung'') differs by up to a factor 4
in the most recent calculations~\cite{kaptari,kaptari2,mosel}.

On the other hand, the different contributions of short-lived sources are not
easy to separate, as, in principle, they have to be treated in a
coherent approach, which is done usually in quantum mechanical
calculations including the interferences, 
using {\it e.g.} One-Boson Exchange (OBE) models. Such calculations for the
process $NN \to NN e^+e^-$ have already been done in~\cite{early_papers}
and more recently in refs.~\cite{kaptari,kaptari2,mosel}. 

Our goal is to present the methods and their applicability to
the HADES $pp/pn$ data
in order to have a simulation tool at our disposal
to be sensitive to additional sources going beyond the
$\Delta$ Dalitz decay contribution.
Due to restricted phase space coverage and efficiency, a flexible
simulation tool, 
which is capable to make direct comparison of model predictions with
data is particularly useful in this respect.  Here we 
describe an extended version of the event generator Pluto~\cite{pluto} 
which is the
standard simulation tool for the HADES experiments. 

The calculations mentioned above ensure a coherent treatment
of the $NN \to NN e^+e^-$ reaction, including graphs involving 
nucleons, $\Delta$'s or higher resonances and fulfill gauge invariance.
As the graph involving the $\Delta$ Dalitz decay process is expected to be dominant
in the $pp$ reaction and still important in the $pn$ reaction, a separate treatment
of this contribution is useful.
Therefore, one has to consider two mechanisms  for  simulations:
either a full calculation
including properly the interference effects, or the
production via resonances ({\it e.g.} the $\Delta$(1232)) and their
subsequent decay, the so called $\Delta$ Dalitz decay model. 

Following these aims a versatile simulation framework has to be able to: 
\begin{enumerate}
\item convert parameterized (or calculated) differential cross sections
  of the $NN \to NN e^+e^-$ reaction into ``exclusive'' events
  and,
  alternatively,
\item produce di-electrons via resonances ($NN \to \Delta N \to NN e^+e^-$)
  using mass-dependent branching ratios and angular distributions.
\end{enumerate}
For each of these two methods different descriptions should be compared.
These new developments are useful wherever simulations of
this kind (in $NN$ as well as in future $\pi N$ experiments)
have to be done in the context of the interpretation of HADES data.

The main goal of this report is to describe a standardized method to
incorporate calculations for $NN \to NN e^+ e^- $ reactions 
wherever available and 
subsequently to compare them 
to experimental data using an open-source and adaptable
package, and to demonstrate how simulations of the $dp/pp$ reaction
may be performed to provide a valuable tool for interpreting
the HADES data~\cite{next_prl}.
Our paper is organized as follows.
In section 2, we outline the numerical implementation, the software
framework and how the simulation of both types have been done. 
The simulation results are discussed in section 3.
Our summary can be found in section 4.
In appendix A, we explain the models included
to describe the $\Delta$ Dalitz decay and the used form factors.  

\section{Numerical realization}

\subsection{Pluto framework}

The simulations which are presented here have been elaborated within
the context of the Pluto framework~\cite{pluto} originally intended to
be used for experiment proposals.  The Pluto package is entirely based
on ROOT~\cite{root} and steers the event production with very little
overhead by using so-called ``macros'' which are - within the ROOT
framework - based on the C++ language.

After the set-up procedure, the event loop is called $N_{\rm ev}$
times which creates the momenta of all involved particles (and masses
in the case of unstable ones) event-by-event. Subsequently, each event
is usually filtered with the detector acceptance or fed into a full
digitization package like Geant~\cite{geant} 
(not part of our package).  The $pp$
(or $pn$) system enters - in our case - as a ``seed object'' into
the decay chain with a given center of momentum (c.m.) total energy
and momentum. Cocktails - which are the incoherent sum of different
reaction channels - can be generated as well.

Recently, the Pluto package was re-designed in order to introduce a
more modular, object-oriented structure, thereby making additions such
as new particles, new decays of resonances and new algorithms up to
modules for entire changes (plug-ins) easily
applicable~\cite{pluto_chep}.

\subsection{The \boldmath$\Delta$ mass shape}\label{mass_sampling}

One method of event-based simulations is to set up a reaction of
consecutive decays, like $pp \to p\Delta^+ \to pp \gamma^*_{ee} \to pp
e^+e^-$. Hence, this means that the $\Delta$ mass shape (and
mass-dependent branching ratios) must be known prior to event
sampling.

How does one usually generate $N_{\rm ev}$ events for the Dalitz decay $\Delta \to N
\gamma^*_{ee}$? In the first step, the $\Delta$ mass
shape is sampled, {\it i.e.}, for each event $i$, a mass $m_\Delta^{(i)}$ is
assigned. 

Following the usual ansatz (see {\it e.g.}~\cite{cite_9_teis}) we use the
relativistic form of the Breit Wigner distribution:
\begin{equation}\label{eqn:breit_wigner}
g^\Delta(m_\Delta) = A \frac{m^2_\Delta \Gamma^{\rm tot}(m_\Delta)} 
{(M^2_\Delta - m^2_\Delta)^2 + m_\Delta^2 (\Gamma^{\rm tot}(m_\Delta))^2}
\end{equation}
where $m_\Delta$ denotes the actual energy (resonance mass), and
$M_\Delta$ is the static pole mass of the resonance. The
mass-dependent width is the sum of the partial widths
\begin{equation}\label{eqn:width_sum}
\Gamma^{\rm tot}(m_\Delta) = \sum^N_{\rm k} \Gamma^{\rm k}(m_\Delta)
\end{equation}
with $N$ the number of decay modes. The factor $A$ is chosen
such that the integral is statistically normalized, $\int dm_\Delta\
g^\Delta(m_\Delta) = 1$, {\it i.e.},
eq.~(\ref{eqn:width_sum}) leads to the following condition for the mass-dependent  branching
ratio for each decay mode $k$
\begin{equation}\label{eqn:br}
  b^{\rm k}(m_\Delta)=
  \frac{\Gamma^{\rm k}(m_\Delta)}{\Gamma^{\rm tot}(m_\Delta)}.
\end{equation}

The width for the dominating hadronic decays $\Delta \to N \pi$ is
derived from a well-known ansatz~\cite{cite_9_teis,cite_7_wolf,cite_8_monitz}:
\begin{eqnarray}\label{eqn:width_monitz}
  \nonumber \Gamma^{\Delta \to N \pi}(m_\Delta) &=&
  \frac{M_\Delta}{m_\Delta} \left(\frac{q^{\pi}
  (m_\Delta)}{q^{\pi}(M_{\Delta})} \right)
  ^{3} \\ && \times \left( \frac{ \nu
  (m_\Delta)}{\nu(M_{\Delta})} \right)
  \Gamma^{\Delta \to N \pi}.
\end{eqnarray}
The dependence on the two decay products with masses $m_N$ and $m_\pi$
enters via the terms $q^{\pi}(m_\Delta)$ and
$q^{\pi}(M_{\Delta})$, namely the momentum of one out of the two
decay products in the rest frame of the parent resonance.  We follow
ref.~\cite{cite_9_teis,cite_8_monitz} which uses for the resonance the
cutoff parameterization
\begin{equation}\label{eqn:cutoff_monitz}
  \nu(m_\Delta)=\frac{\beta^2}{\beta^2+(q^{\pi}(m_\Delta))^2}
\end{equation}
with the parameter $\beta$=300~MeV.

To simulate the Dalitz decay, the mass-dependent
branching ratio has to be taken into account as
\begin{equation}\label{eqn:partial_breit_wigner}
  g^{\Delta \to N
    e^+e^-} (m_\Delta) = b^{\Delta \to N
    e^+e^-}(m_\Delta) g^\Delta(m_\Delta).
\end{equation}
Thus, the partial decay width 
\begin{equation}\label{eqn:delta_dalitz_decay_width_int}
  \Gamma^{\Delta \to N e^+e^-}(m_\Delta) = \int dm_{ee}
  \frac{d\Gamma^{\rm \Delta\to Ne^+e^-}_{m_\Delta}}{dm_{ee}}
\end{equation}
has to be used in the numerator of
eq. (\ref{eqn:breit_wigner}) which is calculated 
by integrating eq.~(\ref{eqn:delta_dalitz_decay_width_int}) 
with a small step size for
each $m_\Delta$ mass bin, respectively.

\begin{figure}
\begin{center}
\resizebox{0.45\columnwidth}{!}{%
  \includegraphics{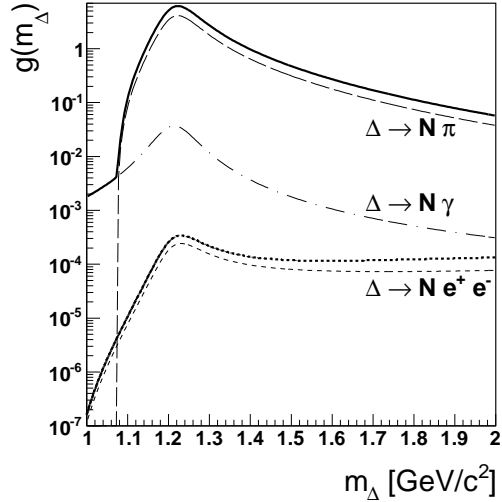}
}
  \caption{ Free spectral shape $g^\Delta$ as a function of $m_\Delta$
    of $\Delta^+(1232)$ (solid line) compared to the distribution
    functions for dedicated decay states. Short dashed line: $\Delta^+
    (1232)\to e^+e^-p$ (ref.~\cite{kriv} with ``photon point'' magnetic form factor), long
    dashed line: $\Delta^+ (1232) \to \pi^0+p$ and dashed-dotted line:
    $\Delta^+ (1232) \to \gamma +p$. In addition, the effect of the two component quark-model transition form factor
    of ref.~\cite{iachello} is
    indicated by the short dotted line.  }\label{fig:delta_mass}
\end{center}
\end{figure}

The differential $\Delta$ Dalitz decay width $d\Gamma/dm_{ee}$ 
depends
on three transition form factors, as explained
in more detail in the appendix.  In particular, we compare 
the effect of using constant  form factors, with same 
values as for a real photon emission (``photon-point'' form factors) with a
two-component model from Iachello and Wan~\cite{iachello} which is in
line with the vector-meson dominance (VMD) model.
This leads to different mass shapes
$g^{\Delta \to N e^+e^-}(m_\Delta)$ which is exhibited in Fig.~\ref{fig:delta_mass}.

One can see that the
mass-dependent branching ratio decreases with mass. This feature is an
important issue for the di-electron production as large
contributions of the higher-mass tail of the $\Delta$ resonance and
the mass-dependence in the $\Delta$ Dalitz branching ratios affect not
only the di-electron invariant-mass spectral shape but also the
di-electron yield, compared to the hadronic channels.  Therefore, the
mass-dependent branching ratio must be considered even if the effect
might be suppressed in part by the limited phase space.

How can an event generator take that behavior into account for the mass sampling?  An
elegant way by using the weighting method is explained in the
next subsections.

\subsection{Weight-based method}\label{simple}

Our first step is to bring the spectrum
(represented by a histogram) onto an absolute scale such that one can
assign each of its bins to a differential cross section.  
Therefore, total cross section models for elementary
$NN$ collisions have been implemented.

For the reactions discussed here, two major contributions (beside the
bremsstrahlung) have been added, which is the $\Delta$
production~\cite{cite_9_teis} 
(we assume $\sigma_{pp \to p \Delta ^+} = \frac{3}{2} \sigma_{pp \to pp \pi ^0}$) 
and the close-to-threshold $\eta$
production~\cite{moskal,calen,calen_d_eta} (the latter one is needed
for the $pn$ case).  
In
order to obtain these cross sections independent of the
number of events, first a default weight of $1/N_{\rm ev}$
is applied to the $NN$ seed object.  In the decay algorithms, where
mass and momentum sampling takes place, the weights of all attached
distributions are multiplied with the default weight of the
parent particle to an event weight $W_i$. In particular,
parameterizations for the total cross section can be included.  This
leads - in a 1-step decay as defined above - to the simple relation

\begin{equation}\label{w_sum}
  \sigma^{NN \to X Z} =
  \sum^{N_{\rm ev}}_i W_i =
  \frac{1}{N_{\rm ev}}
  \sum_{i}^{N_{\rm ev}} W^{NN \to X Z},
\end{equation}
where $X$ is an unstable particle ({\it e.g.} $\pi^0$, $\eta$) and $Z$ stands for the
remaining particles.  Eq.~(\ref{w_sum}) is the basic definition to be used in
all weighted Pluto simulations: The integral (sum) of the resulting histogram
weights represents the total (or partial) cross section of an exclusive
reaction. 
Therefore, the simulated spectra can be compared directly to the normalized 
experimental data.
If we now extend this definition to a two-step process, where a decay
$X \to ab$ follows, the weight of particle $X$ has to be folded with the
mass-dependent weight of the consecutive decay model, which will be described in
Sec.~\ref{sec:model_dg_dm}.

In the simplest case we can consider that production and decay of $X$
are independent, which is the case for the $\eta$ as well as for the $\pi^0$ 
as they have a comparable small
width. In particular, by using a sampling model (returning a random event from
the known distributions and momenta of $a,b$) this weight is the static
branching ratio $W^{X \to ab}=b^{X \to ab}$ yielding

\begin{equation}\label{w_sum2}
  \sigma^{NN \to a b Z} = \frac{1}{N_{\rm ev}}
  \sum_{i}^{N_{\rm ev}} W^{NN \to X Z}  b^{X \to ab}.
\end{equation}

\subsection{Flat di-electron generator}

Very often experiments are concerned with regions of the phase space, where a
small number of events is expected compared to the overall number
of a given process. This is clearly the case for the electromagnetic
Dalitz decays, where the di-lepton yield spans many orders of
magnitude with a high differential cross section for the low-mass
pairs, whereas the high-mass pairs have a much lower cross section
$d\sigma/dm_{ee}$.  Obviously, a large number of events have to be
sampled before an acceptable number in the high-mass region has been
collected.  On the other hand, the Monte-Carlo simulations presented
here need an adequate statistics for the high-mass region of
interest.

The solution used here is that sampling is done using a flat di-lepton
distribution first. 
Then, a weight is calculated
using the same physics decay model,
which was employed for the sampling in Sec.~\ref{mass_sampling}.  This
means, the decay weight $W^{X \to ab}_i$ changes from event to event,
depending on the values $m_i=m_{ee}^{(i)}$.

By extending eq.~(\ref{w_sum2}) we get for the Dalitz decays
of the pseudo scalar mesons

\begin{eqnarray}\label{w_pseudo}
  \sigma^{NN \to NN e^+e^- \gamma}  
  & = & \frac{1}{N_{\rm ev}} \sum_{i}^{N_{\rm ev}} W^{NN \to NN  (\eta, \pi^0)}  
  \\ 
  & 
  \times &
  S^{(\eta, \pi^0) \to e^+e^-\gamma} W^{(\eta, \pi^0) \to e^+e^-\gamma}(m_{ee}^{(i)})   \nonumber
\end{eqnarray}
with $W^{(\eta, \pi^0) \to e^+e^-\gamma}(m_{ee})$
as the differential cross section from~\cite{landsberg}.
The normalization factor
\begin{equation}\label{eqn:def_s}
  S^{(\eta, \pi^0) \to e^+e^-\gamma} = b^{(\eta, \pi^0) \to e^+e^-\gamma}/
  \overline{W^{(\eta, \pi^0) \to e^+e^-\gamma}}
\end{equation}
is used to normalize the spectrum to the selected branching ratio and
thereby correct for the fact that more events are created in the phase
space region with small probability, where the model weight is small
compared to the region usually containing a large number of
events. Here, the average decay weight
\begin{eqnarray}
  \overline{W^{(\eta, \pi^0) \to e^+e^-\gamma}} & =&   \frac{1}{N ^\prime_{\rm ev} } 
  \sum_{i=1}^{N_{\rm ev} ^\prime}  W_i^{(\eta, \pi^0) \to e^+e^-\gamma}, \\
  & & N ^\prime_{\rm ev} = N_{\rm ev} +  N_{\rm pre}
\end{eqnarray}
is first calculated for a selected number of events ($N_{\rm pre}=1000$ turned out
to be sufficient to avoid artefacts) and then adjusted within the
running event loop.

For broad resonances, as discussed in Sec.~\ref{mass_sampling}, the
branching ratio should be an outcome, and no precondition of the
calculation.  The same argument can be brought forward for the
calculations from refs.~\cite{kaptari,mosel}.  Integrating these
distributions is difficult since the underlying calculations are only
presented for invariant di-electron masses larger then
50-100~MeV/c$^2$~\cite{kaptari,mosel}. Thus, the partial cross section
can not be calculated correctly and added to the data base; at least
it would require some extrapolation.  However, this can be avoided, as
shown in the following.

\subsection{Models returning \boldmath$d \sigma/dm$}

As an application for a simulation without an explicit
re-normalization to a fixed branching ratio or total cross section,
the above-mentioned weighting has been exploited because the
calculations from refs.~\cite{kaptari,mosel} provide the differential
cross section $d \sigma / dm_{ee}$ already done on an absolute
scale.  Aiming for a comparison of the $\Delta$ Dalitz decay with the
resonant $N \Delta$ terms (method 1), we define as a model weight 
$W^{N \Delta}(m_{ee}^{(i)})= d \sigma
/ dm_{ee}$ parameterized as described below.
It is evident that the same method can be used for the full (coherent) 
differential cross section and the quasi-elastic term as well, just be 
replacing $W^{N \Delta}(m_{ee}^{(i)})$ in the generator by 
 $W^{\rm full}(m_{ee}^{(i)})$ and $W^{\rm ela}(m_{ee}^{(i)})$, respectively.

In such a case it is more convenient to use the function $\frac{d
  \sigma }{dm_{ee}}$ directly with the flat di-electron distribution
generator, as in the previous example, but without the intermediate step
of production and decay.  As the event loop generates a row of
$N_{\rm ev}$ values $m_i=m_{ee}^{(i)}$ the cross section is represented
 naturally by the Monte-Carlo integration method as

\begin{eqnarray}\label{eqn:monte_carlo}
  \nonumber \sigma^{pp \overset{N \Delta}{\longrightarrow} pp e^+e^-} & = & \int
  \frac{ d \sigma }
       { d m_{ee}} dm_{ee} \\ & \approx & \frac{1}{N_{\rm ev}}
  \sum_{i=1}^{N_{\rm ev}} W^{N \Delta}(m^{(i)}_{ee}) \cdot \Delta m_{ee},
\end{eqnarray}
where $\Delta m_{ee}$ is the kinematic range of the di-electron
generator (changing also event-by-event) which is provided by the
model. It is an advantage that the normalization factor $S$ is not
needed.

The function $W^{N \Delta}$ required here has been obtained by
digitizing the curves provided by ref.~\cite{kaptari} and using a
parameterization of the form

\begin{eqnarray}\label{eqn:obe_param}
  & & W^{N\Delta}(m_{ee}) \nonumber  =
  \frac{d\sigma}{dm_{ee}} (m_{ee})\\ & & =
  \frac{ (m_{\rm max} -m_{ee})^{P_3}  }{P_2 \exp ({P_0 m_{ee} + P_1}m_{ee}^2)}
\end{eqnarray}
with
\begin{equation}\label{eqn:obe_param2}
  m_{\rm max} = \sqrt{(m_{N2} + m_{N1}) ^2 +2m_{N2} T_{\rm kin}} 
  - (m_{N1} + m_{N2})
\end{equation}
as the kinematic limit 
with $N1$ as the beam nucleon. The parameters $P_i$ fitted to the curves in
ref.~\cite{kaptari} are polynomials $P_i(T_{\rm kin}) = a_i^0 \cdot a_i^1
T_{\rm kin} \cdot a_i^2 T_{\rm kin}^2 $.  The function $d \sigma / dm_{ee}$ of ref.~\cite{mosel} was
directly supplied by the authors~\cite{shyam} for a fixed kinetic beam
energy of $T_{\rm kin}=$1.25~GeV, and moreover of $T_{\rm kin}=$1~GeV
and $T_{\rm kin}=$1.5~GeV for the $pn$ case.

\subsection{Models returning \boldmath$d \Gamma/d m$}
\label{sec:model_dg_dm}

Let us continue with an alternative (method 2), where the $\Delta$
production is followed by the Dalitz decay.
Pluto treats this process
in two steps, leaving out the last and uncritical decay $\gamma^*_{ee}
\to e^+e^-$. The weight $W^{\Delta \to N
  \gamma^*_{ee}}(m_\Delta,m_{ee})$ is now a function of two masses.
Taking eqs.~(\ref{eqn:br},\ref{eqn:delta_dalitz_decay_width_int}) the
mass-dependent branching ratio is obtained by

\begin{equation}\label{eqn:delta_weight}
  b^{\Delta \to N e^+e^-} (m_\Delta)
  = \int dm_{ee} \frac{d \Gamma^{\Delta \to N e^+e^-}_{m_{\Delta}} 
    (m_{ee})}{dm_{ee}} \frac{1}{ \Gamma^{\rm (tot)} (m_\Delta)}.
\end{equation}

Similar to eq.~(\ref{eqn:monte_carlo}), 
the di-electron generator represents
the Monte-Carlo integration method of the model
 
\begin{equation}\label{eqn:delta_weight_mod}
W^{\Delta \to N
  \gamma^*_{ee}}(m_\Delta,m_{ee}) = \frac{d \Gamma^{\Delta \to N e^+e^-}_{m_{\Delta}} 
    (m_{ee})}{dm_{ee}}.
\end{equation}
Therefore the mass-dependent branching ratio obtained as
\begin{equation}\label{eqn:delta_weight2}
  b^{\Delta \to N e^+e^-} (m_\Delta) \approx
  \frac{1}{N_{\rm ev}}
  \sum_{i=1}^{N_{\rm ev}} \frac{W^{\Delta \to N
  \gamma^*_{ee}}(m_\Delta,m_{ee}) \cdot \Delta m_{ee}}
      { \Gamma^{\rm (tot)} (m_\Delta)}
\end{equation}
is already considered, which means that the $\Delta$ mass shape has to
be sampled in the first step with the pure function $g^\Delta(m)$
without using condition~(\ref{eqn:partial_breit_wigner}).
This is ensured by the Pluto framework automatically.  The effect of
the mass dependence, as discussed in Sec.~\ref{mass_sampling} becomes
now visible: The fraction of the di-electron events, compared to the
hadronic channels,
\begin{equation}\label{eqn:energy_dep_br}
  b(T_{\rm kin})=\frac{\sigma^{NN \to N\Delta \to NN e^+e^-}}{\sigma^{NN \to N\Delta}},
\end{equation}
is in our case with $b(T_{\rm kin}=$1.25~GeV)=4.96~$\cdot 10^{-5}$
significantly larger then the static branching ratio, which is 
$b^{\Delta \to N ee}=4.19 \cdot 10^{-5}$ 
(see \ref{sec:constant_ff}).
This means that the mass-dependent branching ratio is an important
feature which can not be neglected.

\section{Simulation}

\subsection{\boldmath$\Delta$ production}

From the experimental point of view the consideration of known angular
distributions is crucial. This is in particular true for the emission
of the $\Delta$ resonance in the c.m. frame which affects the
direction of both protons. The impact on the experimentally measured
data comes due to the fact that the detection of the proton
could be a trigger requirement or, explicitly, enters the analysis
of a \mbox{(semi-)}exclusive channel, {\it e.g.} $pp \to Xpe^+e^-$. Such effects are
often integrated out in calculations.

In the case of the $\Delta$ production, we follow the one-pion exchange model of
ref.~\cite{cite_13_dmitriev}, which is in excellent agreement with the
data and in detail described in ref.~\cite{pluto}. This adds a strong
forward-backward peaking of the polar angle of the nucleons with
respect to the beam axis, an effect which is considered in the
two-step Dalitz decay simulations. In the OBE simulations
virtual photons are generated with the differential cross sections
taken from the coherent calculation like a hypothetical ``on-mass''
shell particle with a given mass $m_{ee}$.  Due to missing information
the momentum and energy is sampled assuming a
three body phase space decay of the $NN$ seed object sitting at rest
in the center-of-mass with $m=m_{Np}$. Consequently, the
virtual photons are emitted isotropically in the OBE, but
not in the $\Delta$ simulation.

\begin{figure*}
\begin{center}
\resizebox{0.4\textwidth}{!}{%
  \includegraphics{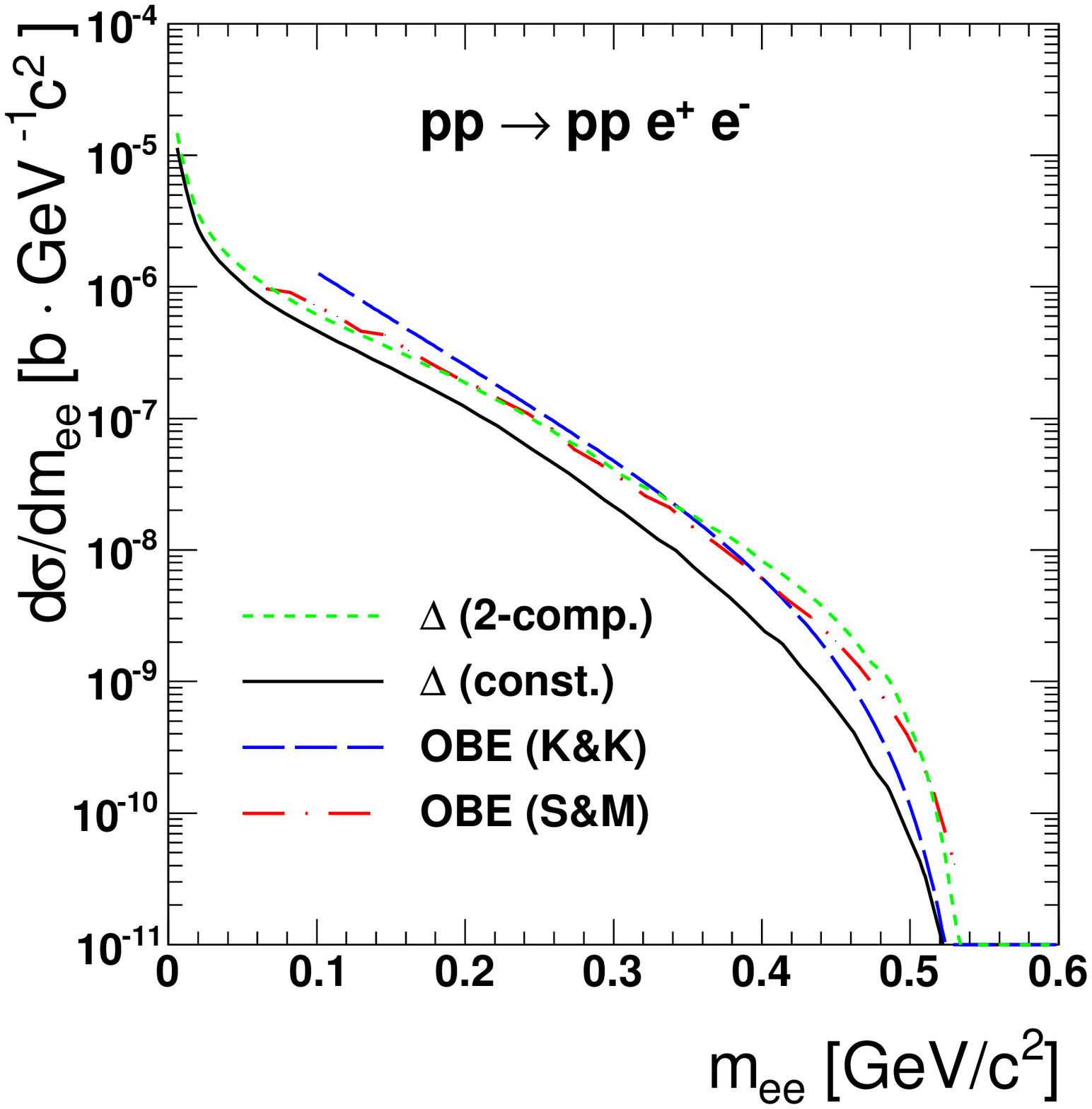} }
\resizebox{0.4\textwidth}{!}{%
  \includegraphics{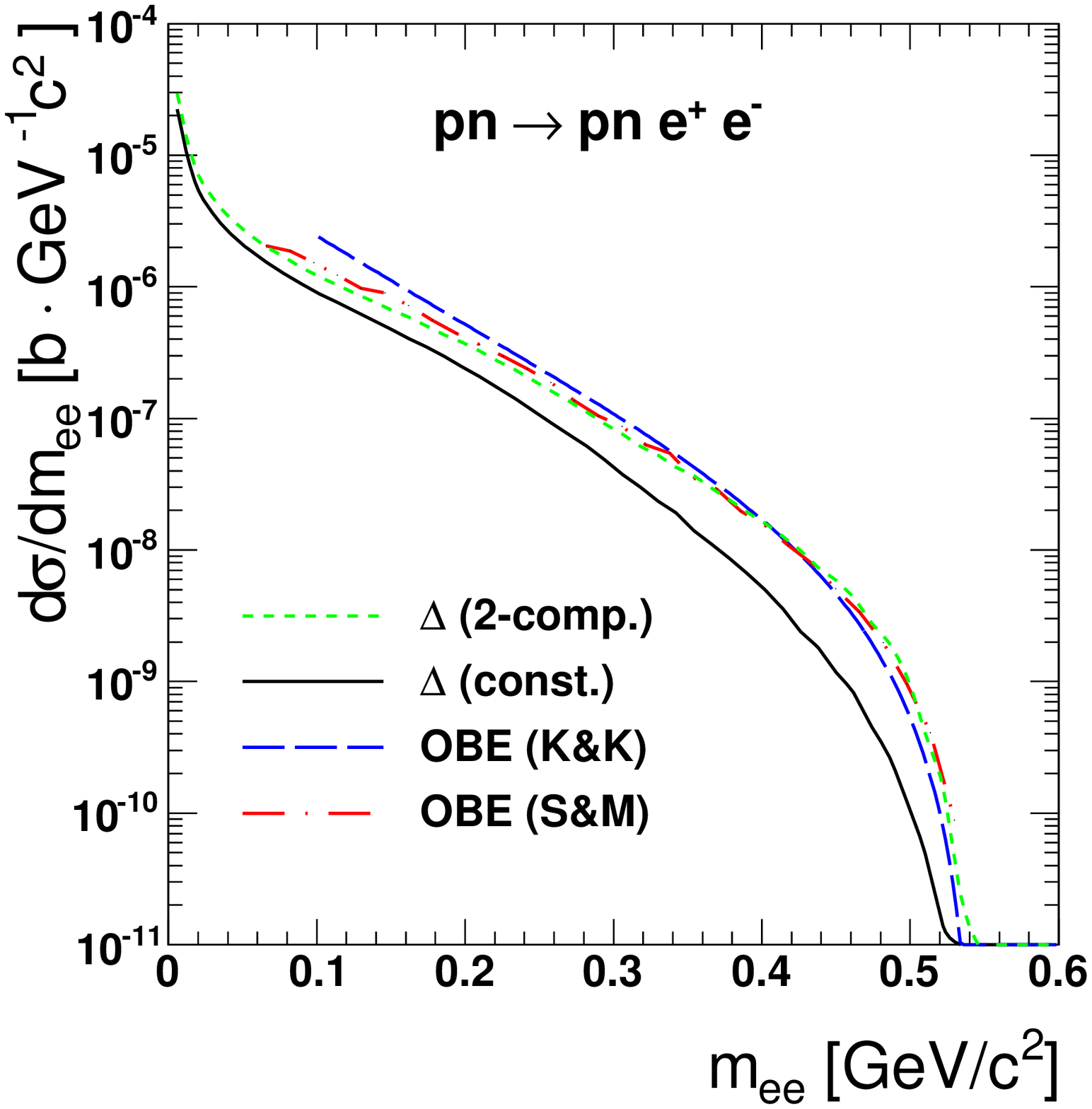}
}
  \caption{
    (color online)
    Resonant di-electron production in $Np \overset{N
      \Delta}{\longrightarrow} Np e^+e^-$ reactions at $T_{\rm kin}$=1.25~GeV.
    {Left:} $pp$ reaction, {Right:} $np$ reaction.  The  solid black
    curve and short-dashed  (green) curves are the result of the full Pluto simulation
    using the mass-dependent $\Delta$ width, its production cross section and
    the $d \Gamma(m_\Delta) / dm_{ee}$ description
    from~\cite{kriv} with constant transition form factors 
    (labeled with ``const.''), 
    the short-dashed (green) was obtained with the two componant form factor~\cite{iachello} 
    (labeled with ``2-comp.'') 
    in addition.
    The long-dashed (blue) curve
    is the 
    calculation from~\cite{kaptari}, whereas the (red) dot-dashed curve is based on
    ref.~\cite{mosel} and provided by ref.~\cite{shyam}.
  }\label{fig:delta_dilepton}
\end{center}
\end{figure*}

\subsection{Different descriptions of the \boldmath$\Delta$ Dalitz decay}
\label{sec:dd}

After having all needed pieces at our disposal the difference between
the OBE calculations~\cite{kaptari,mosel} and the calculations
from~\cite{kriv} (either with the photon point form factors or the
two componant quark model option) can be studied in
a quantitative way.  Fig.~\ref{fig:delta_dilepton} shows the Pluto
simulation for the processes $pp \to p\Delta^+ \to pp e^+e^-$ (left)
and $pn \to N\Delta^{+,0} \to pn e^+e^-$ (right) for the two
descriptions mentioned above together with the resonant term $Np
\overset{N \Delta}{\longrightarrow} Np e^+e^-$ from the OBE
calculation~\cite{kaptari}.  
The two OBE calculations use the same 
$N-\Delta$ transition form factors, but the latter are different than the
``photon point'' form factors used in the two step Dalitz decay  model.
The 
effect on the di-electron spectrum is, however, expected to be lower then 15\%.

It is obvious that the disagreement is much larger. The
OBE calculation of ref.~\cite{kaptari} is larger by a factor of 2-4
than the production via a ``free'' $\Delta$ using the mass-dependent
branching ratio. The latter method is a crude approach,
taking into account only one graph 
neglecting interferences and anti-symmetrization effects.
It is, however, frequently used
in transport code calculations.  

The VMD calculation comes closer to
the OBE model, but undershoots it at low masses as well.
Surprisingly, we do not come to the same conclusion as
ref.~\cite{mosel}, where the spectrum of the two-step Dalitz decay
model has a very steep slope. Qualitatively,
our model is much closer to the OBE
calculations of ref.~\cite{kaptari,mosel} 
and, taking the VMD form factor into account, 
lies almost
on top the curves from  ref.~\cite{mosel}.
This clearly needs experimental confirmation and
further theoretical studies.

\subsection{Final state interaction}

In all near-threshold reactions, the final state interactions  (FSI) 
may influence strongly both the total cross section as well as the population of
the phase space. The first effect is already included in the Delta Dalitz two-step
simulation as they use measured data.
For the OBE calculations the factorization~\cite{sib,kaempfer2}
\begin{equation}\label{eqn:fsi_fac}
  W^{\rm final} = W^{\rm fsi} \times W^{N \Delta, \rm full, ela} 
\end{equation}
has been implemented.  This is controlled via a switch which means a
factor $W^{\rm fsi}$ is attached exploiting the inverse of the Jost
function
\begin{equation}\label{eqn:fsi}
  W^{\rm fsi} = \left( \frac{1}{J(k)} \right)^2 =\left(  \frac{k + {\rm i}
  \alpha}{k + {\rm i} \beta} \right)^2
\end{equation}
with 
\begin{equation}\label{eqn:fsi2}
  \alpha, \beta = \frac{1}{r_0}
  \left(
    \sqrt{1-\frac{2r_0}{a_0}} \pm 1
  \right),
\end{equation}
and $k$ as the relative momentum of the two nucleons using
the effective radii and scattering lengths $a_0 = -7.8098$~fm and $r_0
= 2.767$~fm for the $pp$ case, and $a_0 = -23.768$~fm and $r_0 =
2.75$~fm for the $pn$ case~\cite{kaempfer2}. We use this
description to be compatible with the calculation done
in~\cite{kaptari}; other functions could be implemented as well.


\subsection{Nucleon momentum distribution in the deuteron}

\begin{figure}
\begin{center}
\resizebox{0.45\columnwidth}{!}{%
  \includegraphics{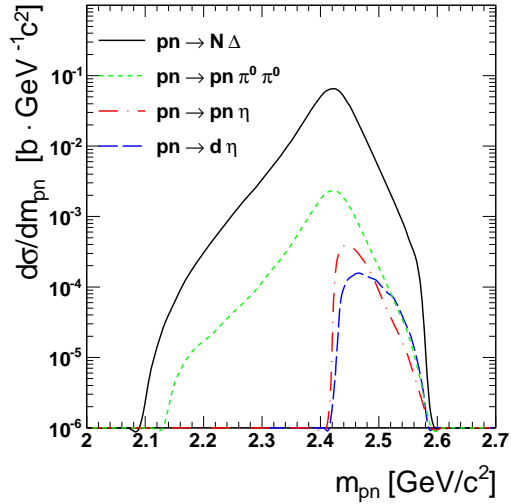}
}
  \caption{
     (color online) Differential cross section as a function of the invariant mass $m_{pn}$
  in the quasi-free $pn$ reaction taking into account the deuteron momentum distribution.
  Solid curve: $\Delta^{+,0}$ production, short dashed (green) curve: $pn \to
  pn \pi^0 \pi^0$ (constant $\sigma$ = 0.1~mb assumed), long-dashed (blue) curve: $pn
  \to pn \eta$, dot-dashed (red) curve: $pn
  \to d \eta$.
  }\label{fig:dp_eta_cross}
\end{center}
\end{figure}

For the $dp$ reaction, the nucleon momentum distribution in the
deuteron 
has to be taken into
account, because the effective neutron momentum may have a big impact
at larger di-electron masses~\cite{kaptari2}.  This is done by Pluto
in a two-step process.  In the first step, the off-shell mass of the
participant
\begin{equation}\label{eqn:fermi}
  m_{\rm part} ^2 = 
  m_d ^2 + m_p ^2  - 2m_d \sqrt{
  m_p ^2 +p_{\rm Deut}^2  }
\end{equation}
is determined by the parameterized wave function $p_{\rm Deut}$ from
ref.~\cite{benz}. Along with the second reaction particle (in our case
the target $p$ at rest) this off-shell particle forms the $pn$
composite with a total c.m. energy $m_{pn}$ of the quasi-free
reaction.  In a second step the bremsstrahlung model calculates the
energy
\begin{equation}\label{eqn:fermi2}
  T_{\rm kin} = \frac{m_{pn} ^2 - m_p^2 - m_n^2 - 2m_pm_n}{2m_n}
\end{equation}
of the proton in the neutron rest frame.
Here, we use the invariant mass $m_{pn}$ to get the total
cross section $\sigma(m_{pn})$ using the on-shell neutron mass $m_n$, 
which is also
the approach in ref.~\cite{ulrike}. 
It is important to note that, as the the di-electron cross section for 
the $pn$ reaction was parameterized as a
function of the kinetic proton energy, the actual $T_{\rm kin}$ reflects the
proton quasi kinetic energy even in the case of a deuteron beam and a proton
target.

The consequence of the above-mentioned momentum distribution 
is that the $dp$ reaction results in a
``smeared'' $pn$ reaction c.m.  energy and enables thus sub-threshold
$\eta$ production~\cite{moskal,calen,calen_d_eta}.  
 The formation of $pn$ composites
below any threshold of the final state (as {\it e.g.} for the $pn \to pn
\eta$ threshold) is rejected while counting the number of rejected events 
to keep the proper normalization.

Fig.~\ref{fig:dp_eta_cross} shows the different channels which are combined
later on in the final di-electron cocktails.

\subsection{\boldmath$pn/pp$ ratio}

\begin{figure}
\begin{center}
\resizebox{0.45\columnwidth}{!}{%
  \includegraphics{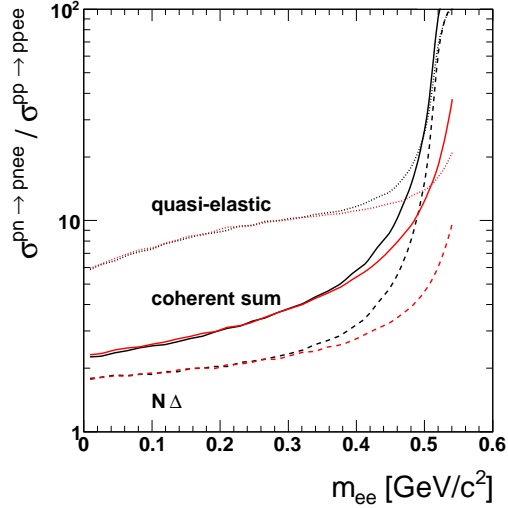}
}
  \caption{
    Ratio of $\sigma^{pn \to pn ee} / \sigma^{pp \to pp ee}$ as a function
    of $m_{ee}$ using the OBE
    calculation~\cite{kaptari} with the resonant $N\Delta$ term (dashed
    curves), the quasi-elastic term (dotted curves) and the full calculation
    (solid curves), for a beam energy of 1.25~GeV/u. The grey (online: red) 
    curves are obtained with the pure $pn$ 
    reaction, whereas the black curves include the momentum distribution of the deuteron.
  }\label{fig:ratio}
\end{center}
\end{figure}

In the following, the iso-spin dependence  of the
bremsstrahlung is discussed. By using the same methods for the OBE calculations
for the resonant term, the quasi-elastic term and the full calculation,
respectively, and dividing the simulation results for $pn$ and $pp$ the iso-spin
dependence can be studied. Fig.~\ref{fig:ratio} shows this ratio for the
different contributions and the coherent sum, 
optionally including the momentum distribution of the deuteron. 
The simulation indicates
that the influence of the momentum distribution for masses smaller than 0.4~GeV/c$^2$ is
negligible. 

\begin{figure*}
  \begin{center}
    \resizebox{0.4\textwidth}{!}{%
      \includegraphics{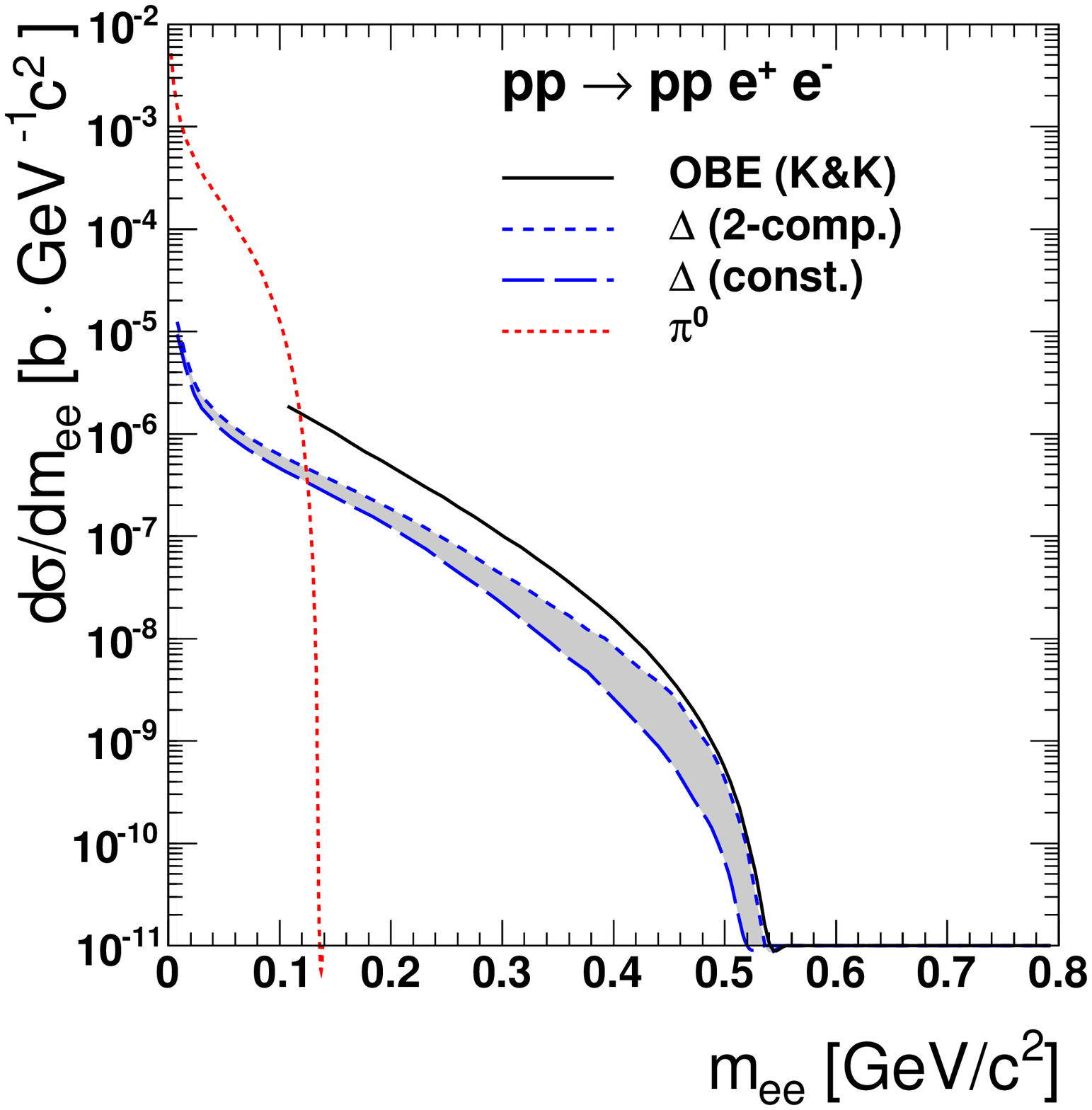}
    }
    \resizebox{0.4\textwidth}{!}{%
      \includegraphics{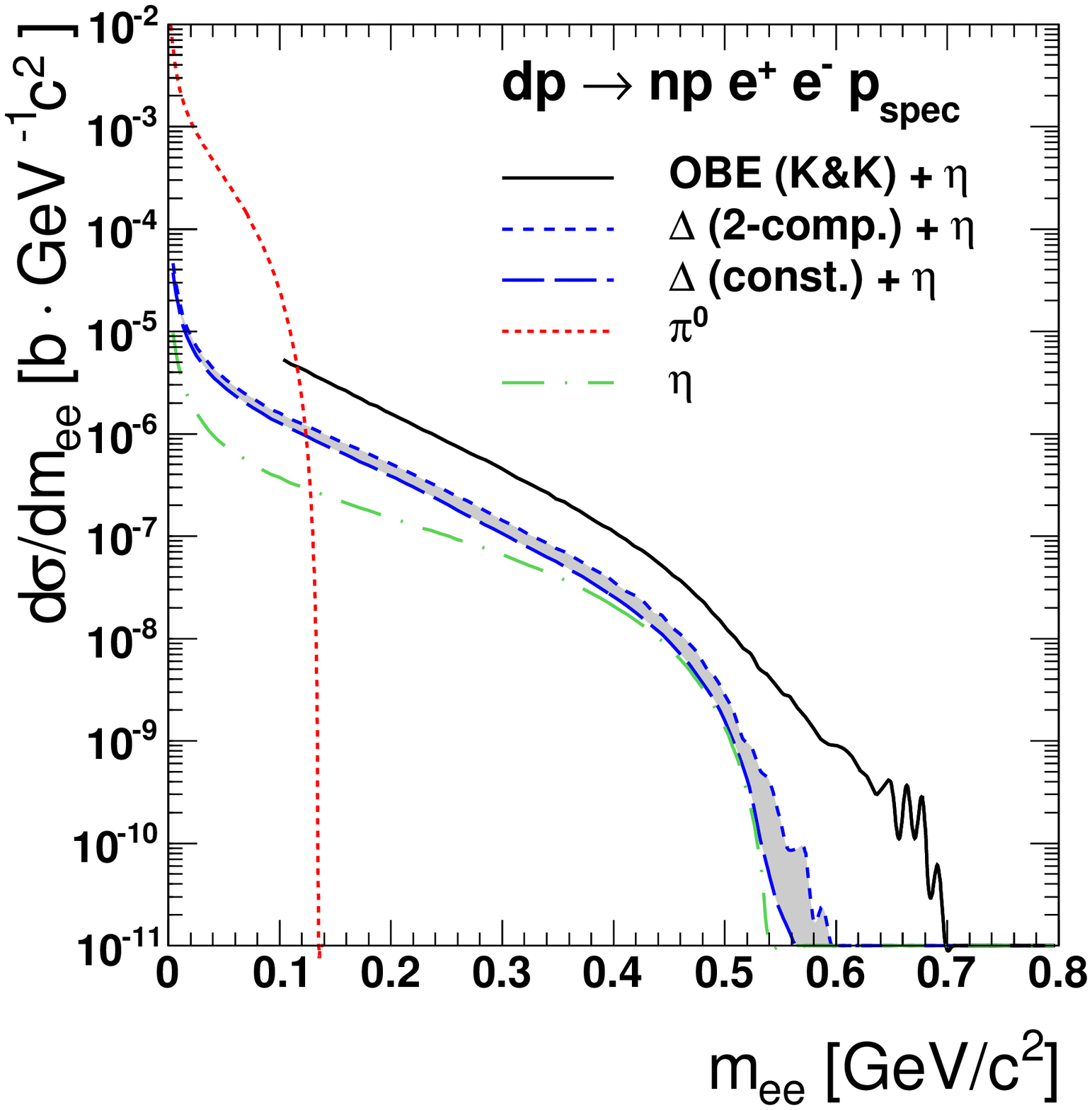}
    }
    \caption{ 
       (color online)
      Cocktail simulation of the differential cross section
      of $d\sigma / d m_{ee}$   
      for the $pp$ reaction at 1.25~GeV (left) and
      the $dp$ reaction at 1.25~GeV/u (right).
      Long dashed (blue) curve: $\Delta$ Dalitz decay using constant
      transition form factors, short dashed (blue) curve: same with 
      the two component quark model from~\cite{iachello,new_iach},
      solid black curve: full coherent OBE
      calculation from~\cite{kaptari}, dot-dashed (red) curve: $\pi^0$ Dalitz decay,
      dotted (green) curve: $\eta$ Dalitz decay ($dp$ only).
      In the $dp$ case, the $\eta$ contribution haas been added to the short-lived components.
    }\label{fig:cocktail_comparison}
  \end{center}
\end{figure*}

\subsection{Cocktail simulation}

In the context of the new HADES data, the simulation has to be done with a
full cocktail calculation. Here, all channels contributing to di-electron
production at given energy have to be included. The main source of the di-electrons are the
$\pi^0$ Dalitz decays. The production of $\pi^0$ mesons is done within the
$\Delta$ resonance model assuming that all $\pi^0$'s are created via $\Delta$. The
production of $\eta$ mesons is also included for the $dp$ case.


Fig.~\ref{fig:cocktail_comparison} shows the possible contributions
for the di-electron production, based on (i) the model from
refs.~\cite{kaptari,kaptari2}, and (ii) on the $\Delta$ production and
subsequent
Dalitz decay. The difference between these two approaches is clearly
visible. Moreover, the Dalitz decay
is used with the form factor model from Iachello and
Wan~\cite{iachello}.

We expect that the HADES data will be sensitive to differences
in the various
descriptions. However, the simulation done with the Pluto framework has to be
filtered first with the acceptance of the HADES spectrometer. One limitation
is here that the OBE calculations~\cite{kaptari,kaptari2,mosel} 
provide only the invariant mass spectra, but
do not show predictions for angular distributions. In the $\Delta$ Dalitz decay
model from Pluto, however, these effects are taken into account using existing
data~\cite{pluto}. We therefore would like  to suggest that future theoretical
investigations should consider and present these distributions
as this would be extremely helpful for the 
disentanglement as well as understanding of the aforementioned effects, 
specifically concerning the HADES data.

\section{Summary}

In summary we have presented the details of an extension of the
previous Pluto framework able to incorporate important descriptions for the
production of low-mass di-electrons in
elementary collisions such as $pp$ and quasi-free $pn$. Several models
based on a free $\Delta$ Dalitz decay and full quantum mechanical
calculations have been included so far and are ready for the
comparison with the upcoming HADES data~\cite{next_prl}.  As the intermediate
excitation of $\Delta$ is important for the di-electron production, we
have described in detail the corresponding models.

\section*{Acknowledgments}

We greatly appreciate fruitful discussion with F. Iachello about the
two-componant model and thank for the permission to use his calculations
in our event generator.  Furthermore, the authors would like to thank
R. Shyam for the support of a new set of calculations compatible to
the HADES energies.  Interesting discussions with U. Mosel are greatly
appreciated.

This work was supported by the Hessian LOEWE initiative through the
Helmholtz International Center for FAIR (HIC for FAIR) and by the
Helmholtz Alliance EMMI ``Extremes of Density and Temperature: Cosmic
Matter in the Laboratory'', and BMBF 06DR135.

\begin{appendix}

\section{ \boldmath$\Delta$ Dalitz decay models}

\subsection{Differential decay width of the \boldmath$\Delta$ Dalitz decay}

In this report, for computing the $\Delta(1232)$ Dalitz decay we use
the prescription of Ref.~\cite{wolf} 

\begin{equation}\label{eqn:wolf}
\frac{d\Gamma^{\rm \Delta\to Ne^+e^-}_{m_\Delta}}{dm_{ee}}(m_{ee})
=\frac{2\alpha}{3\pi m_{ee}} \Gamma^{\Delta \to N
  \gamma^*}_{m_\Delta}(m_{\gamma ^*}{ \equiv m_{ee}}).
\end{equation}
with $\alpha = 1/137$ as the fine structure constant.
The decay process $\Delta \to N \gamma ^* $ consists in three
independent amplitudes, which can be calculated unambiguously from the
electromagnetic vertex. However, as stressed in~\cite{kriv},
inconsistent formula for the differential decay width of this process
can be found in the literature. We tackled the calculation by
ourselves, using the magnetic dipole, electric quadrupole and Coulomb
quadrupole covariants~\cite{jones}, and could confirm the expression
of~\cite{kriv}, as repeated below:
\begin{eqnarray}\label{eqn:delta_dalitz_decay_kriv}
  \nonumber \Gamma^{\Delta \to N\gamma^*}_{m_\Delta}(m_{\gamma^*}) & =
  & \left( G^{\Delta \to N\gamma^*}_{m_\Delta}(m_{\gamma^*}) \right)^2
  \\ & & \nonumber \times \frac{\alpha}{16} \frac{(m_\Delta +
    m_N)^2}{m_\Delta^3m_N^2} {\sqrt{y_{_{\scriptstyle +}} y_{_{\scriptstyle -}}^3}}, \\ y_\pm & =& (m_\Delta
  \pm m_N)^2-m_{ee}^2,
\end{eqnarray}
where the index $N$ refers to the produced nucleon, $e$ is the electron
charge, and $G^{\Delta \to N\gamma^*}_{m_\Delta}(m_{\gamma^*})$ { depends on the N-$\Delta$} 
electromagnetic transition form factors as

\begin{eqnarray}\label{eqn:form_factor}
  \nonumber
\left(G^{\Delta \to N\gamma^*}_{m_\Delta}(m_{\gamma^*})\right)^2  & = &
  \left|G_M^2(m_{\gamma^*})\right| + 3 \left|G_E^2(m_{\gamma^*})\right| \\ & & +
  \frac{m^2_{\gamma^*}}{2m_\Delta^2} \left|G_C^2(m_{\gamma^*})\right|
\end{eqnarray}
where G$_M(m_{\gamma^*})$, G$_E(m_{\gamma^*})$, G$_C(m_{\gamma^*})$
are the magnetic, electric and Coulomb $N-\Delta$
transition form factors, respectively, which will be discussed in the next section
of the appendix. Note that eq.~(\ref{eqn:delta_dalitz_decay_kriv})
implies a normalization of the form factors as in~\cite{jones}, since
isospin factors are included in the numerical factors. We could also
check the validity of the expressions derived in~\cite{zet}, where the
amplitudes are calculated with a different, but equivalent set of
covariants, with corresponding form factors.
We use eqn.~(\ref{eqn:delta_dalitz_decay_kriv}) throughout.

\subsection{Electromagnetic { \boldmath$N-\Delta$ transition form factors} }
\label{sec:em}

The electromagnetic $N-\Delta$ transition form factors are analytical
functions of the squared four-momentum transfers $q^{2}$ at the
$N-\Delta$ vertex. Pion electroproduction and photo production
experiments allow to determine these formfactors in the space-like
region ($q^2 \leq$0)~\cite{pasca}. In the $\Delta$ Dalitz decay process,
due to the positive four-momentum transfer squared
($q^2 = m_{\gamma^*}^2 >{ 4m_{ee}^2}$), the time-like region is probed, where  only the limit at $q^2=0$ is known experimentally. An
additional difficulty hails from the fact that the form factors,
which are real in the space-like region, get an imaginary part in the
time-like region. 

Therefore, two options can be
chosen to compensate the lack of experimental information on these
observables and have been implemented for Pluto simulations.

\subsubsection{Constant $N-\Delta$ transition form factors}\label{sec:constant_ff}

In this option, which
  is based on the smallness of the squared four-momentum transfer
  $q^2$ in the $\Delta$  Dalitz decay process, the form-factors are
  given the values G$_M$=3.0, G$_E$=0, G$_C$=0. This choice is
  consistent with the precise measurements of the electric and
  magnetic form factors in pion photoproduction experiments~\cite{pasca}
  and with the very small contribution of the Coulomb term
  in eq.~(\ref{eqn:form_factor}) and provides in addition the correct radiative
  decay width $\Gamma ^{\Delta \rightarrow N \gamma^*}=0.66$~MeV. The resulting
  branching ratio at the
  pole mass of $b^{\Delta \to N ee}=4.19 \cdot 10^{-5}$ is remarkably
  consistent to the photon decay branching ratio times the fine
  structure constant $\alpha$ which would result in $b^{\Delta \to N
  ee}=4.01 \cdot 10^{-5}$.

\subsubsection{Two-component  quark  model}

The alternative is to use a model for the $N-\Delta$ transition form
factors. This model should in principle satisfy the analyticity
properties in the complex $q^2$ plane, as well as the asymptotic
behavior predicted by QCD sum rules, while reproducing the existing
data measured in the space-like region. At first, the photon-point
value provides the normalization of the whole function.  As an example
of such models, the two-component quark model,
described in the following,
 has been implemented in
the Pluto event generator.

\begin{figure}
\begin{center}
\resizebox{0.45\columnwidth}{!}{%
  \includegraphics{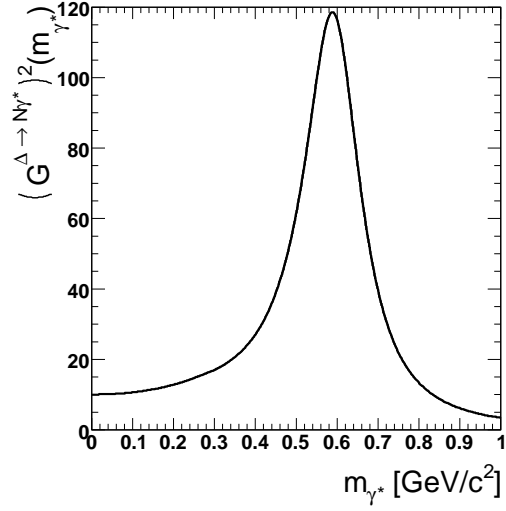}
}
  \caption{ 
    Form factor model taken from~\cite{iachello,new_iach}.
  }\label{fig:iachello_ff}
\end{center}
\end{figure}

The picture behind this transition form-factor model is the overlay of
an intrinsic q$^3$ structure and a meson cloud which couples to the
virtual photon via vector mesons~\cite{iachello_jackson}.  The model
allows an analytical continuation of the function from space-like to
time-like region~\cite{iachello_timelike}, inducing a phase, and could
be successfully applied to the description of elastic space-like and
time-like nucleon form-factor measurements.  The formalism has been
recently extended to calculate also baryonic transition form factors
in a unified way~\cite{iachello}, these new developments have been
tested on the space-like $N-\Delta$  transition form factors
measurements. We use here the simplest version of the model, which
assumes isospin symmetry, and therefore considers only the magnetic
$N-\Delta$  transition form factor.  The intrinsic 3-quark structure is
described as:
\begin{equation}\label{eqn:form_factor_iachello2}
  g(q^2)=\frac{1}{(1-a^2e^{i\theta}q^2)^2}
\end{equation}
and the overall expression for the time-like $N-\Delta$  transition form factor is:  
\begin{eqnarray}
  \nonumber
    G_{M}(q^{2}) & = &
    \mu_{p}\left(\frac{4}{3\sqrt{2}}\right)\sqrt{\frac{2m_{N}m_{\Delta}} 
      {m_{\Delta}^2+m_N^2}}g(q^2) \\ & 
    \times  & (\beta^{'} +\beta F_\rho(q^2) ),
   \label{eq:Iachello}
\end{eqnarray}
where $\mu_{p} = 2.793$ 
is the proton magnetic moment and
$\beta$ and $\beta^{'}$ are the constants for the
coupling to the quark core and to the meson-cloud respectively.  In
the case of the $N-\Delta$ transition, only the $\rho$ meson
contributes to the latter contribution, due to isospin conservation
and the corresponding $q^2$-dependence is given by F$_\rho(q^2)$, as:

\begin{eqnarray}\label{eqn:form_factor_iachello3}
F_\rho(q^2)=\frac{m_{\rho}^{2}+8\Gamma_{\rho}m_{\pi}/\pi}{m_{\rho}^{2}-q^{2}+4m_{\pi}(1-x)\Gamma_{\rho}(\alpha(x)-i\gamma(x))},
 \end{eqnarray}
where we have introduced ${\displaystyle x=q^2/4m_{\pi}^2}$ and~\cite{new_iach}
 \begin{eqnarray}
  \left.  \begin{array}{cc}
 {\displaystyle
   \alpha\left(x\right)= \frac{2}{\pi} \sqrt{\frac{x-1}{x}}
   \ln{\left(\sqrt{x-1}+\sqrt{x}\right)} }    
   & \\[-3mm]
    & \\
   \hspace{5mm}  &    \\ 
    & \\[-3mm]  
 {\displaystyle    
  \gamma\left(x\right) = \sqrt{\frac{x-1}{x}}       }
          \end{array}
  \right\}    
  \ \mathrm{if } \  x \, > 1
  \label{eqn:form_factor_iachello4}
  \end{eqnarray}
  
   \begin{eqnarray}
 \left.  \begin{array}{cc} 
 {\displaystyle
 \alpha \left(x\right)=\sqrt{\frac{1-x}{x}}\left[1-\frac{2}{\pi} \cot^{-1}{\sqrt{\frac{x}{1-x}} }\right] } 
  & \\[-3mm]
    & \\
   \hspace{1mm}  &    \\ 
    & \\[-3mm]  
 {\displaystyle    
  \gamma\left(x\right) = 0       }
          \end{array}
  \hspace{-2mm} \right\}  
 \label{eqn:form_factor_iachello4a}  
 \ \mathrm{if } \  x  \, < 1.  \hspace{3mm}
   \end{eqnarray}

\begin{figure}
\begin{center}
\resizebox{0.45\columnwidth}{!}{%
  \includegraphics{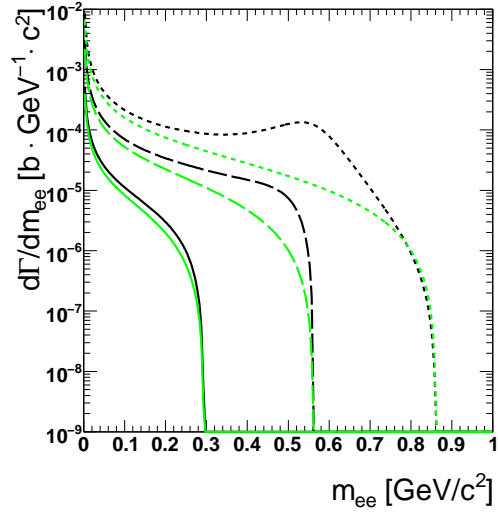}
}
  \caption{ 
    (color online)
    The distribution
    $\frac{d\Gamma^{\rm \Delta\to
        Ne^+e^-}_{m_\Delta}(m_{ee})}{dm_{ee}}$ 
    as a function of $m_{ee}$
    for 3 different $\Delta$
    masses. Solid curves: pole mass $m_\Delta=1.232$~GeV, long dashed curves:
    $m_\Delta=1.5$~GeV and short dashed curves: $m_\Delta=1.8$~GeV. The
    description from~\cite{kriv} is applied. The upper set of black curves has been
    calculated with the two-componant quark model from ref.~\cite{iachello} whereas the lower set
    of grey (online: green) have been obtained with the constant transition amplitudes
    as described in the text.  }\label{fig:delta_width}
\end{center}
\end{figure}

In eq.~\ref{eqn:form_factor_iachello3}, the values $m_\rho$=765~MeV and 
$\Gamma_\rho$=112~MeV are used different from the physical values 
due to the form of parameterization.
The value of the parameters a=0.29 GeV$^{-2}$, $\theta$=53$^{\circ}$,
$\beta$=1.2147 and $\beta$'=0.004 results from fits of the model
predictions to the available experimental information as discussed
above. The resulting distribution of form-factor values
(see fig.~\ref{fig:iachello_ff}) shows a broad peak centered around
$\sqrt{q^2}=m_{ee} \sim 0.6 m_\rho^2$. Due to the small value of the $\beta'$
parameter, the contribution of the coupling to the quark core in this
model is negligible up to $q^2$= 5~(GeV/c)$^2$, the dominant feature
of the model in the kinematic range probed by the Dalitz decay process
is therefore the vector dominance.

Fig.~\ref{fig:delta_width} finally shows the distribution { $d\Gamma^{\rm
    \Delta\to Ne^+e^-}_{m_\Delta}/dm_{ee}$} for the two form factor
models. The results clearly exhibit a rising decay width for larger
$\Delta$ masses. { In a proton-proton collision at 1.25 GeV incident energy, 
the mass of the produced baryonic resonance is limited to 1.48 GeV/c$^{2}$, the latter effect will nevertheless affects the shape 
 of }
the di-electron mass spectrum shown in Sec.~\ref{sec:dd}.
The $\Delta$ Dalitz decay branching ratio,
defined at the pole mass is however mainly determined by the values of
the form factor at very low $q^2$. The branching ratio obtained using
the two-component quark model form factor is about 10\% larger then with
the constant ``photon-point'' value, as discussed in~\ref{sec:constant_ff}.
This derives from the fact that the parameters of the model are fitted to a set
of data over a large $q^2$ range, which results in a slightly too
high value for the magnetic form factor at $q^2 = 0$.

One should note that, in the semi-classical description, this would
correspond to the production of an on-shell $\rho$ meson, which should
therefore
not be added then as an independent contribution when enabling the two-component quark model.

\end{appendix}

%

\end{document}